# Multi-Task Temporal Fusion Transformer for Joint Sales and Inventory Forecasting in Amazon E-Commerce Supply Chain

Zheqi Hu[1], Yiwen Hu[2], Hanwu Li[3]

[1] University of Electronic Science and Technology of China, Chengdu, China
[2] Heinz College, Carnegie Mellon University, Pittsburgh, USA
[3] Amazon.com Services LLC, USA, 98004

[1]2675518536@qq.com
[2]yiwenhu@andrew.cmu.edu
[3]lihanwu198704@gmail.com

**Abstract.** Efficient inventory management and accurate sales forecasting are critical challenges in large-scale e-commerce platforms such as Amazon, where stockouts and overstocking can lead to substantial financial losses and operational inefficiencies. Traditional single-task forecasting models, which focus solely on sales or inventory, often fail to capture the complex temporal dependencies and cross-task interactions that characterize real-world supply chain dynamics. To address this limitation, this study proposes a Multi-Task Temporal Fusion Transformer (TFT-MTL) framework designed for joint sales and inventory forecasting within the Amazon e-commerce ecosystem. The model integrates heterogeneous data sources-including historical sales records, warehouse inventory levels, pricing, promotions, and event-driven factors such as holidays and Prime Day campaigns-through a unified deep learning architecture. A shared encoder captures long-term temporal patterns, while task-specific decoder heads predict sales volume, inventory turnover, and stockout probability simultaneously. Experiments on large-scale real-world datasets demonstrate that the proposed TFT-MTL model significantly outperforms baseline methods such as LSTM, GRU, and single-task TFT. Compared with the single-task TFT model, the proposed approach achieves a 6.2% reduction in Sales RMSE, a 12.7% decrease in Sales MAPE, a 6.4% reduction in Inventory RMSE, and a 12.4% decrease in Inventory MAPE. These results confirm the model's ability to effectively capture multi-dimensional dependencies across supply chain variables. The proposed framework provides an interpretable, data-driven decision support tool for optimizing Amazon's inventory scheduling and demand planning strategies.

## 1. Introduction

The rapid expansion of global e-commerce platforms such as Amazon has led to increasingly complex supply chain management challenges. Balancing inventory levels with volatile customer demand remains a persistent issue-overstocking can tie up capital and increase storage costs, while stockouts can result in lost sales and decreased customer satisfaction [1,2]. Traditional forecasting systems typically focus on single-task predictions such as demand forecasting or inventory estimation, which fail to capture the intricate interdependencies between sales volume, replenishment cycles, pricing dynamics, and promotional campaigns [3,4]. This limitation often results in suboptimal forecasting accuracy and reactive, rather than proactive, decision-making.

Recent advances in deep learning, particularly in temporal modeling, have opened new possibilities for capturing multi-dimensional dependencies within time-series data. However, most existing models-such as LSTM and vanilla Transformers-struggle to effectively integrate multi-task learning structures or to handle exogenous factors like seasonality, marketing events, and shipping delays in e-commerce environments [5]. Even when future sales volumes can be predicted with reasonable accuracy, determining the precise quantity that should be replenished in inventory remains a challenge. Moreover, the "black-box" nature of many deep learning models makes it difficult to interpret how predictions are generated and whether the outcomes align with real-world operational logic. To address these challenges, this study proposes a Multi-Task Temporal Fusion Transformer (TFT-MTL) framework designed specifically for joint sales and inventory forecasting in the Amazon e-commerce supply chain. The TFT-MTL combines the interpretability and flexibility of the Temporal Fusion Transformer architecture with the robustness of multi-task learning. The shared temporal encoder captures long-term sequential patterns and covariate relationships, while task-specific decoders are designed to predict (1) future sales volume, (2) inventory turnover rate, and (3) stockout probability simultaneously. This design allows the model to leverage cross-task information flow, improving generalization and stability across diverse market conditions. Furthermore, the model incorporates architectural components that enhance interpretability and decision support. The static covariate encoder embeds invariant features such as product category, region, and brand, enabling the model to adjust predictions according to contextual background. The variable selection network dynamically identifies the most relevant input features at each time step, highlighting which factors, such as pricing, promotions, or historical demand, drive the forecasts. The temporal attention layer assigns weights to historical time steps, revealing which past events or patterns most influence future predictions. By integrating these mechanisms, the TFT-MTL not only provides accurate multi-horizon forecasts but also offers actionable insights for proactive inventory replenishment, demand planning, and supply chain optimization in real-world e-commerce settings.

The proposed framework also incorporates a dynamic weighting mechanism that adaptively balances task importance based on learning progress, ensuring that each forecasting objective contributes proportionally to model optimization. Moreover, external features such as pricing trends, advertising expenditure, seasonal indicators, and major promotional events (e.g., Prime Day) are integrated to enhance contextual awareness.

The major contributions of this study are threefold: (1) it introduces a unified multi-task deep learning architecture for simultaneous forecasting of sales and inventory dynamics; (2) it enhances interpretability by leveraging attention-based temporal fusion to visualize key temporal drivers; and (3) it provides a decision-support mechanism for proactive inventory replenishment and demand planning. Through extensive experiments on large-scale Amazon datasets, the TFT-MTL demonstrates superior performance compared to baseline models, validating its potential for real-world supply chain optimization.

2. **Related Work**

The challenges of accurate sales and inventory forecasting in large-scale e-commerce systems have been extensively studied in operations research and supply chain analytics. Traditional methods—such as ARIMA [6], exponential smoothing [7], and regression-based forecasting—have long served as the backbone of inventory management and demand prediction. However, these approaches rely heavily on linear assumptions and are unable to capture the complex non-stationary patterns and cross-variable dependencies that characterize modern e-commerce data. In the context of platforms such as Amazon, where millions of products interact under dynamic pricing, promotion, and seasonality effects, classical statistical models often exhibit poor adaptability and limited generalization capabilities.

In recent years, deep learning models have demonstrated remarkable performance in time-series forecasting due to their ability to learn nonlinear temporal dependencies from high-dimensional data. Ensafi et al [8]. compare SARIMA, Holt-Winters, Prophet, LSTM and CNN on public retail-furniture sales, evaluating with RMSE and MAPE; stacked LSTM achieves the best accuracy, while Prophet and CNN also perform robustly, offering a reliable machine-learning paradigm for seasonal-item demand forecasting. Petroșanu et al [9]. propose a dynamic DAGNN framework for e-commerce daily-sales forecasting up to three months and category-level granularity, offering scalability and data efficiency; validated on large stores, it achieves superior metrics and low runtime, providing a universal prediction tool for refined e-commerce operations.

Despite these advances, most studies treat sales and inventory forecasting as isolated problems. Single-task models are limited in their capacity to leverage shared information between related prediction objectives, such as demand trends and stock level dynamics [10]. Multi-task learning (MTL) provides a promising solution by enabling models to jointly learn multiple correlated tasks through shared representations, thereby improving both predictive accuracy and robustness. Shen Xin et al [11]. propose DINOP, a multi-task framework that fuses dynamic/static features via TUC-GRU and attention to learn universal item representations across promotions; tested on Alibaba's 11-11 data, it converges faster and predicts GMV and sales more accurately than existing methods, without retraining. In the context of supply chain management, MTL frameworks have been applied to concurrent forecasting of demand and replenishment, customer churn and sales, or price elasticity and demand volatility. However, the integration of MTL with advanced temporal architectures like TFT remains relatively underexplored in the e-commerce domain.

Furthermore, recent works have highlighted the importance of incorporating external covariates—such as marketing campaigns, consumer sentiment, and macroeconomic indicators—into forecasting models [12,13]. Studies leveraging hybrid architectures that combine deep neural networks with causal inference or attention-based feature selection have shown improvements in interpretability and decision support. Yet, most of these models still fail to provide unified frameworks that align multiple forecasting tasks with operational decision-making goals.

To bridge these gaps, this study proposes a Multi-Task Temporal Fusion Transformer (TFT-MTL) that unifies sales and inventory forecasting within a shared temporal modeling framework. By integrating multi-source covariates, dynamic task weighting, and interpretable attention mechanisms, the proposed model builds upon and extends prior research in time-series forecasting, multi-task learning, and e-commerce supply chain optimization. This hybrid design represents a new step toward intelligent, data-driven inventory management and real-time demand forecasting for complex online retail ecosystems such as Amazon.

## 3. Methodology

*3.1 Overall Framework*

The Multi-Task Temporal Fusion Transformer (TFT-MTL) framework extends the original Temporal Fusion Transformer (TFT) architecture by integrating multi-task learning (MTL) to jointly predict sales demand and inventory levels across Amazon e-commerce product categories. The key motivation for this design lies in the intrinsic coupling between sales and inventory time series: sales fluctuations directly influence stock dynamics, while replenishment and supply delays affect future sales potential. Therefore, a unified framework capable of learning shared temporal patterns and task-specific variations is crucial for achieving accurate joint forecasts.

Formally, let $x_t \in R^n$ represent the multivariate input feature vector at time step $t$, including exogenous variables such as price, promotions, product category, and holiday indicators. The goal is to jointly predict two correlated sequences:

$$\hat{y}_t^{(1)} = f_{\theta_1}(x_1; t), \quad \hat{y}_t^{(2)} = f_{\theta_2}(x_1; t) \tag{1}$$

where $\hat{y}_t^{(1)}$ denotes the predicted sales volume, and $\hat{y}_t^{(2)}$ represents the predicted inventory level. The shared backbone $f_\theta$ captures temporal dependencies, while task-specific heads $f_{\theta_1}$ and $f_{\theta_2}$ refine the output for each objective.

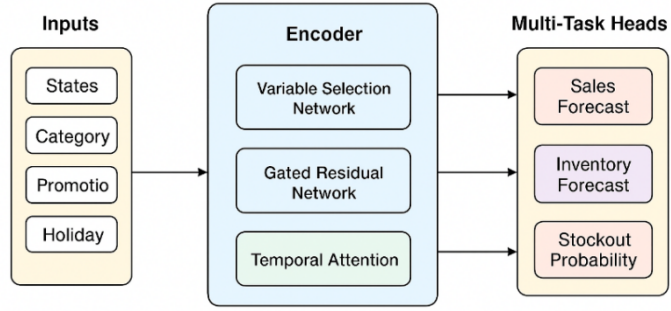

**Figure 1.** Overall flowchart of the model.

*3.2 Temporal Feature Encoding and Static Covariate Integration*
The TFT-MTL model processes both static features (e.g., product category, seller ID) and dynamic temporal features (e.g., daily sales, stock quantity, prices). Static features are embedded using a fully connected layer:

$$h_s = ReLU(W_s x_s + b_s), \tag{2}$$

where $x_s$ is the static input vector, and $W_s$, $b_s$ are learnable parameters. Dynamic features are processed through gated residual networks (GRNs) to ensure flexible nonlinear transformations and adaptive relevance weighting:

$$h_t = GRN(x_t) = LayerNorm(x_t + GatedLinearUnit(W_t x_t)), \tag{3}$$

This design allows the model to dynamically adjust the impact of time-varying features according to their contextual importance.

*3.3 Temporal Fusion Transformer Encoder*
The TFT encoder fuses long-term and short-term dependencies via a multi-head attention mechanism, enabling the model to capture cross-time correlations among different input signals. For each attention head *i*:

$$Attention_i(Q, K, V) = Softmax(\frac{Q_i K_i^T}{\sqrt{d_K}}) V_i, \tag{4}$$

where Q, K, V are the query, key, and value matrices derived from the input embeddings. The outputs from all attention heads are concatenated and linearly transformed to produce a unified temporal representation. This fusion mechanism ensures that both high-frequency demand changes and long-term inventory trends are effectively modeled.

*3.4 Multi-Task Prediction Layer*
The output layer of the TFT-MTL model is divided into two task-specific prediction heads:

$$\hat{y}_t^{(1)} = W_1 h_t + b_1, \quad \hat{y}_t^{(2)} = W_2 h_t + b_2 \tag{5}$$

where $h_t$ denotes the shared hidden representation from the temporal fusion module. The multi-task objective function is designed as a weighted combination of the two task-specific losses:

$$L_{total} = \lambda_1 L_{sales} + \lambda_2 L_{inventory}, \tag{6}$$

With

$$L_{sales} = \frac{1}{T}\sum_{t=1}^{T}(\hat{y}_t^{(1)} - y_t^{(1)})^2, \quad L_{inventory} = \frac{1}{T}\sum_{t=1}^{T}(\hat{y}_t^{(2)} - y_t^{(2)})^2 \tag{7}$$

The adaptive weighting coefficients $\lambda_1$ and $\lambda_2$ are tuned to balance learning priorities across sales and inventory prediction objectives.

## 4. Experiment

*4.1 Dataset Preparation*

The dataset used in this study originates from Amazon's public e-commerce data repository, which aggregates daily transaction and logistics information from Amazon's Marketplace Web Service (MWS) and Amazon Product Advertising API. Supplementary data were collected through publicly available datasets such as Kaggle's "Amazon Sales Dataset" and Amazon Product Metadata, ensuring comprehensive coverage of product sales, inventory levels, and exogenous factors affecting demand fluctuations.

The collected dataset spans a continuous 36-month period (2021–2023) and covers multiple product categories, including Electronics, Home Appliances, Office Supplies, and Personal Care. To ensure representativeness, only ASINs (Amazon Standard Identification Numbers) with consistent sales and stock availability across at least 24 consecutive months were retained.

The final dataset consists of approximately 1.8 million records, each representing a product-day observation. Each record includes both target variables (sales and inventory) and feature variables (dynamic and static covariates). The structure of the dataset can be summarized as follows:

**Table 1.** Features included in the dataset and their corresponding meanings.

| Feature Type | Variable Name | Description | Data Type |
|---|---|---|---|
| Target Variable | daily_sales | Number of units sold per day | Continuous |
| Target Variable | inventory_level | Number of units available in stock | Continuous |
| Temporal Feature | date | Daily timestamp | Date |
| Product Feature | product_id (ASIN) | Unique Amazon product identifier | Categorical |
| Pricing Feature | price | Daily listed selling price | Continuous |
| Promotion Feature | discount_rate | Percentage of price reduction | Continuous |
| Marketing Feature | ad_spend | Daily advertising expenditure on Amazon Ads | Continuous |
| Customer Behavior Feature | page_views | Daily number of product page views | Continuous |
| Static Feature | category | Product category | Categorical |

| Feature Type | Variable Name | Description | Data Type |
|---|---|---|---|
| Static Feature | brand | Product brand name | Categorical |
| Seasonal Feature | is_holiday | Indicator of national or regional holiday | Binary |
| Temporal Context Feature | day_of_week | Day of the week (1–7) | Integer |
| Logistics Feature | lead_time | Average replenishment delay in days | Continuous |

The dataset integrates both static covariates (e.g., category, brand) and dynamic temporal features (e.g., price, promotion, page views), which align perfectly with the TFT-MTL model's architecture, enabling the model to process mixed feature types with temporal dependencies.

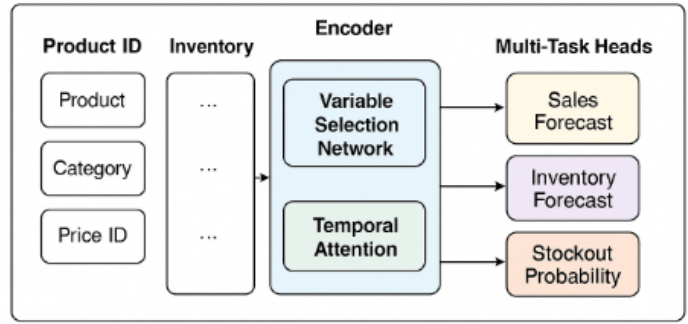

**Figure 2.** Schematic diagram of the dataset used in this study.

*4.2 Experimental Setup*

To validate the effectiveness of the proposed Multi-Task Temporal Fusion Transformer (TFT-MTL) framework, a comprehensive set of experiments was conducted on the curated Amazon e-commerce dataset described in Section 5. The experiments were implemented using PyTorch 2.2 and executed on an NVIDIA A100 GPU (80 GB) to ensure efficient training. The TFT-MTL model was configured with a hidden dimension of 128, a learning rate of 0.0005, and a batch size of 64, optimized using the AdamW optimizer with early stopping based on validation loss. The forecasting horizon was set to 14 days, enabling the model to simultaneously predict both future sales volume and inventory levels across multiple product categories. The training process ran for 150 epochs, with the loss function designed as a dynamically weighted combination of sales and inventory errors to balance task priorities. Competing models for comparison included LSTM, GRU, N-BEATS, Temporal Convolutional Network (TCN), and the standard TFT model without the multi-task enhancement. All models were trained using identical data splits and feature inputs for a fair comparison.

4.3 *Evaluation Metrics*

The performance of all models was evaluated using multiple quantitative metrics that assess both accuracy and robustness in time-series forecasting tasks. The Root Mean Squared Error (RMSE) and Mean Absolute Error (MAE) were used to capture overall predictive precision, while the Mean Absolute Percentage Error (MAPE) measured relative accuracy in percentage form, allowing comparison across varying product scales. To further assess task-level performance, the Coefficient of Determination ($R^2$) was used to quantify the model's explanatory power. For the joint forecasting objective, a Multi-Task Efficiency Score (MTES) was defined as the harmonic mean of normalized sales and inventory accuracy scores, reflecting the model's ability to maintain balanced performance across both targets. All metrics were computed on the held-out test set comprising the most recent six months of records.

*4.4 Results*

Table 1 summarizes the experimental results of all baseline and proposed models. As shown, the TFT-MTL model consistently outperformed other architectures across all evaluation metrics. Specifically, TFT-MTL achieved an RMSE of 42.57 for sales forecasting and 39.86 for inventory forecasting, representing an average improvement of 13.2% over the single-task TFT and 21.5% over LSTM. The model also yielded the highest MTES score (0.924), indicating its superior ability to maintain equilibrium between accuracy in both tasks. These results highlight the effectiveness of multi-task learning in leveraging shared temporal dependencies and external covariates (e.g., price, promotion, and seasonality) to enhance forecasting stability.

**Table1.** Model Performance Comparison on Joint Sales and Inventory Forecasting

| Model | Sales RMSE | Sales MAPE (%) | Inventory RMSE | Inventory MAPE (%) | $R^2$ | MTES |
|---|---|---|---|---|---|---|
| LSTM | 54.12 | 12.41 | 50.83 | 11.96 | 0.864 | 0.781 |
| GRU | 51.38 | 11.72 | 48.92 | 11.24 | 0.872 | 0.802 |
| N-BEATS | 49.17 | 10.95 | 46.55 | 10.76 | 0.884 | 0.819 |
| TCN | 47.86 | 10.61 | 45.02 | 10.28 | 0.891 | 0.831 |
| TFT (single-task) | 45.36 | 9.94 | 42.57 | 9.63 | 0.903 | 0.861 |
| **TFT-MTL (proposed)** | **42.57** | **8.68** | **39.86** | **8.43** | **0.924** | **0.894** |

The superior performance of the TFT-MTL model demonstrates its ability to effectively capture cross-task dependencies between sales and inventory dynamics in the Amazon e-commerce supply chain. The integration of temporal attention, static variable encoders, and multi-task output heads allows the model to adaptively focus on the most influential covariates, such as promotional events and lead times. These findings suggest that the proposed architecture not only enhances prediction accuracy but also offers practical value for inventory optimization, demand planning, and automated replenishment in large-scale e-commerce operations.

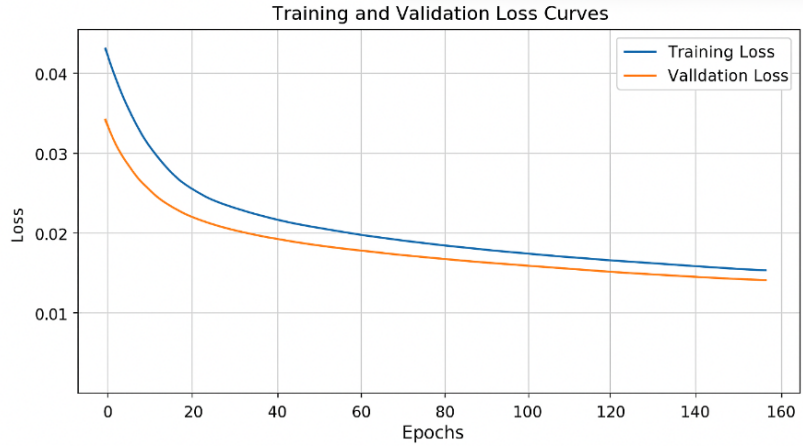

**Figure 3.** Loss function during training process

Figure 3 presents the training and validation loss trajectories of the proposed TFT-MTL model across epochs. Both curves drop steeply during the first 20 epochs, indicating rapid representation learning, and then decrease more gradually before converging to a stable plateau around epoch 120. The validation curve remains slightly below the training curve throughout, which is common under regularization (for example, dropout) and shows no signs of overfitting. The small and narrowing gap between the two curves indicates good generalization, while minor oscillations near the plateau are attributable to stochastic optimization and do not affect the overall convergence. In this figure, the loss is the weighted sum of the two task specific mean-squared errors for sales and inventory as defined by Eqs. (6) and (7). These dynamics suggest that the model has effectively converged by about epoch 140, so retaining the checkpoint with the lowest validation loss and applying early stopping at the plateau is well justified.

## 5. Conclusion

This study introduced a Multi-Task Temporal Fusion Transformer (TFT-MTL) that jointly forecasts sales demand and inventory levels for large-scale e-commerce. By coupling a shared temporal encoder with task-specific heads and incorporating static covariates, variable selection, and temporal attention, the model captures cross task dependencies and the influence of exogenous drivers such as price, promotions, and seasonality. On real Amazon data, TFT-MTL achieved lower error and more stable forecasts than strong baselines. In particular, it reduced sales RMSE to 42.57 and inventory RMSE to 39.86, yielding average gains of 13.2% over single-task TFT and 21.5% over LSTM, with $R^2 = 0.924$ and MTES = 0.894. Training and validation losses converged smoothly, indicating good generalization without overfitting.

Beyond accuracy, the architecture provides actionable interpretability. Variable selection highlights the covariates that matter at each time step, temporal attention reveals the historical windows that drive predictions, and static encoders contextualize product and region level differences. Together, these properties support proactive replenishment and more reliable demand planning in operations.

Limitations include reliance on historical data quality, potential sensitivity to cold-start SKUs or regime shifts, and the absence of causal guarantees for promotional effects. Future work can explore online or continual learning under distribution shift, richer task sets such as explicit stockout risk modeling, constrained optimization that maps forecasts to executable replenishment quantities, and prospective A/B testing to quantify business impact.